\documentclass[sigconf]{acmart}

\copyrightyear{2020} 
\acmYear{2020} 
\setcopyright{acmcopyright}
\acmConference[CEP 2020]{Computing Education Practice 2020}{January 9, 2020}{Durham, United Kingdom}
\acmBooktitle{Computing Education Practice 2020 (CEP 2020), January 9, 2020, Durham, United Kingdom}
\acmPrice{15.00}
\acmDOI{10.1145/3372356.3372358}
\acmISBN{978-1-4503-7729-4/20/01}

\begin{document}

\title{First Year Computer Science Projects at Coventry University}
\subtitle{Activity-led integrative team projects with continuous assessment}

\author{Simon Billings}
\affiliation{%
  \institution{Coventry University}
  \streetaddress{}
  \city{Coventry}
  \state{U.K.}
}
\email{Simon.Billings@coventry.ac.uk}

\author{Matthew England}
\affiliation{%
  \institution{Coventry University}
  \streetaddress{}
  \city{Coventry}
  \state{U.K.}
}
\email{Matthew.England@coventry.ac.uk}

\renewcommand{\shortauthors}{S. Billings and M. England}

\begin{abstract}
We describe the group projects undertaken by first year undergraduate Computer Science students at Coventry University.  These are integrative course projects: designed to bring together the topics from the various modules students take, to apply them as a coherent whole.  They follow an activity-led approach, with students given a loose brief and a lot of freedom in how to develop their project.  

We outline the new regulations at Coventry University which eases the use of such integrative projects.  We then describe our continuous assessment approach: where students earn a weekly mark by demonstrating progress to a teacher as an open presentation to the class.  It involves a degree of self and peer assessment and allows for an assessment of group work that is both fair, and seen to be fair.  It builds attendance, self-study / continuous engagement habits, public speaking / presentation skills, and rewards group members for making meaningful individual contributions. 
\end{abstract}



\maketitle

\section{Our Context}

The authors teach for the undergraduate BSc Computer Science (CS) degree at Coventry University, which has an annual first year intake of around $280$ students.  Our intake is diverse: we make no requirement of prior CS study or programming experience, although some students will have a great deal of this.  

Like most degrees we ensure all students take part in group projects.  We have a second year software engineering module which is a natural place for these, but we also use group projects throughout the first year where they are perhaps less common.  
We place great emphasis on our students obtaining an industrial placement\footnote{A one year secondment from university study where students work full-time in a role certified by their university as providing meaningful experience for the degree topic.} in between Year 2 and 3, for which they typically have to apply mid-way through the second year.  It is thus important that by this point of application they have experiences of team work and leadership that they can draw from in their applications.  This is the main motivation for our extensive use of group projects at Year 1.

We also use these first year group projects to help build a course community, embed good learning habits, and deliver critical information that is not related to a specific module e.g. induction, careers, university procedures, what is plagiarism.  These projects are course-integrative, that is, they are designed to bring together all the different skills and knowledge students are acquiring from their classes so that they can apply them as a coherent whole to a single project.  As described in Section \ref{SEC:Module}, we have been doing this for many years with our success inspiring university regulatory changes to encourage other degree courses to do similar.

In Section \ref{SEC:Groups} we describe how we administer group work and then in Section \ref{SEC:CA} we discus our most recent innovation: the introduction of a continuous form of assessment involving weekly presentations and self-evaluation, to replace the single assessment at the end. 
We finish in Section \ref{SEC:Results} by detailing the positive benefits we have seen through the use of this assessment type, both through qualitative student comments and quantitative data on grades and assessment.

\subsection{Other Recent Innovations at Coventry}

The second author has published at CEP on innovations within the programming curriculum at Coventry: in particular the use of the learning environment Codio \cite{CE19} and the summative testing tool CodeRunner \cite{CE20}.  
Codio provides students with online virtual Linux boxes, which staff equip with guides and tasks with automated marking and feedback.  We adopted it as a response to a low-level of formative feedback provision and uptake (plus rapidly increasing student numbers).  
CodeRunner is a Moodle plugin which we use for summative assessment of programming: it provides an additional Moodle quiz question type where a student's answer is code which is then evaluated against unit tests.  In \cite{CE19, CE20} we describe the benefits, difficulties and student views on our use of these tools.

These innovations are alongside and in support of the projects but do not directly affect them: although students often choose to develop their project code within Codio $-$ for the projects Codio is nothing more than an IDE.

\section{Integrative projects at Coventry}
\label{SEC:Module}

\subsection{First Year Projects 2009/10$-$2014/15}

Integrative course projects in the first year of our computing curriculum date back to 2009/10.  As described in \cite{APHESLLR12} they were initially an attempt to stop early disengagement and improve student retention during the difficult transition from school to university.  Courses ran six week group projects and used activity-led learning.

\emph{Activity-led learning} refers to a learning process where students are presented first with a task and then acquire any skills / knowledge they need in its solution.  It is also referred to as \emph{problem-based learning} and \emph{inquiry-led learning}.  We prefer the name activity-led learning (and use the abbreviation ALL) at it emphasises that our task requires practical activity rather than book-based research.  See the textbook \cite{HLR11} for an overview of ALL in computer science.

\subsection{First Year Projects 2014/15$-$2017/18}

An increase in retention was achieved, leading to the use of ALL projects  throughout our entire first year.  To maintain engagement it was decided that these should now form summative assessment.  
Project sessions were timetabled but they were not formerly constituted as modules.  Instead, they contributed marks to each of the other modules in the first year, where they were combined with marks from assessments dedicated to the topic of that module.  

A separate course project allows students to experience multiple learning styles for each topic.  E.g. the programming modules use a \emph{Computing as Craft}\footnote{An apprentice (student) learns first by observing a master (teacher) and then attempting tasks themselves with feedback from the master or journeymen (PhD Students).} pedagogy \cite{VPL11} with a more didactic teaching style focussing on particular topics in turn.  But students also learn programming via the ALL pedagogy, acquiring additional programming knowledge \emph{as needed} for their projects.  Rather that exclusively using one approach the degree deliberately uses both to recognise that different students react better to different methods\footnote{and different staff teach better with different methods!}.  

There is a similar diversity in assessment: the main modules provide traditional assessment, whose grades are supplemented by the projects. A diversity of assessment types is widely recognised as beneficial\footnote{such diversity is required by our university regulations and external accreditation.} (students excel under different conditions).  The projects efficiently allow for diversity within module (not just course).

However, many university systems did not support the integrative nature of these projects well: necessitating for example manual grade calculations and distribution of grades by email which introduced a risk of human error plus delays in grade finalisation.  There were also regulatory worries $-$ the projects needed to assess large numbers of specific learning outcomes from the different modules, and some staff worried about students passing modules on the basis of project work unrelated to that particular module.  

\subsection{First Year Projects 2018/19 Onwards}

Students starting their degrees at Coventry University in academic year 2018/19 onwards are subject to a new set of regulations with regards to assessment\footnote{Our regulations are available at the following URL with Section 6 covering assessment and 6b the new regulations: \url{https://www.coventry.ac.uk/the-university/key-information/registry/academic-regulations/}}.  One of the key changes was the distinction made between assessment credits and teaching credits.

\subsubsection{CATS Credits}

Like most UK universities we quantify our courses using CATS credits\footnote{\url{https://en.wikipedia.org/wiki/Credit_Accumulation_and_Transfer_Scheme}}.  One credit indicates 10 hours of study (contact time and self-study) with a undergraduate degree made up of 360 credits.  Degrees are then split into modules accordingly: e.g. the standard 120 credit academic year could be delivered as 3 large 40 credit modules or 6 smaller 20 credit modules.  The CATS credits are defined as a measure of study but typically also determine the measure of assessment:  e.g. a 40 credit module is worth twice as much to a degree classification as a 20 credit module).

\subsubsection{Credits at Coventry}  Our new assessment regulations distinguish between \emph{learning credits} (which follow the definition above) and \emph{assessment credits} which determine a student's final degree grade and classification.  Individual modules may have more or less assessment credits than learning credits.  The total numbers of each credit type must be equal over every semester and match the CATS standards.  The main motivation for this is to allow for course integrative assessment only modules to be constituted $-$ working towards the university's strategy of teaching being increasingly course-focussed rather than module-focussed.

\subsubsection{Our New First Year CS} We redesigned our CS degree when moving to the new regulations\footnote{Programme specification is here: \url{https://www.coventry.ac.uk/globalassets/media/documents/registry/course-specs/eec/ug-eec/bsci-computer-science-part-a.pdf}} and in doing so formalised our first year projects as modules with only assessment credits.  Our Year 1 first semester now consists of the following modules:
\begin{description}
\item[4000CEM] Programming and Algorithms
\item[4001CEM] Software Design
\item[4002CEM] Mathematics for Computer Science
\item[4006CEM] ALL Project 1  
\end{description}
and our Year 1 second semester consists of:
\begin{description}
\item[4003CEM] Object Oriented Programming
\item[4004CEM] Computer Architecture and Networks
\item[4005CEM] Database Systems
\item[4007CEM] ALL Project 2
\end{description}
In each list the first three modules have fewer assessment credits than learning credits with the ALL projects containing only assessment credits.  Each project is designed to relate to the other modules in a semester but we now have much greater freedom to focus the projects on particular topics which we want to emphasise.  For example, the first ALL project has a large emphasis on basic programming and building confidence with the standard control structures plus how to work with unfamiliar library code by reading the documentation, but only minimal engagement with mathematics.  This is perfectly acceptable under the regulations.  Similarly, the software design principles learnt in 4001CEM are expected to be used (and are assessed) within the second ALL project, even though these are in different semesters.  

The project modules are constituted not to assess the learning outcomes of the other modules in their semester, but the more general and integrative course learning outcomes.  

At Year 2 and 3 we also use integrative assessment-credit only modules for individual projects; and group projects within traditional modules, but these are not discussed further in this paper.

\section{Group Work}
\label{SEC:Groups}

\subsection{Contact Time}

Our project modules do have some sessions (despite the zero learning credits implying no hours) but these hours are used more for administration.  We use these to form groups, give briefings on the projects, and perform the continuous assessment (next section).  We also use these hours to deliver key information that all students new to university require in context: e.g. on avoiding plagiarism before they write their first essay, and on careers guidance before creating their online portfolio.  There are a few sessions where students can work on their project $-$ here any teacher input is only as a facilitator or moderator of discussions, see e.g. \cite{BV98}.

\subsection{Forming Groups}

In the first semester we form groups semi-randomly via which tables students sat at in the first class.  In the second semester we have experimented with different approaches.  In the past we have formed groups based on attendance data for the first semester, to reflect the most common complaint being when a team member does not show up.  However, this can hamper the ability of students to \emph{turn things around} if they end up in the disengaged group.  

We now try to group students based on programming ability, noting the recent substantial study \cite{BJCS19} which suggests that groups of similar ability allow for better learning outcomes.  We originally did this via quiz grades but now do it via self-selection.  This avoids giving students an external target on which to blame team problems and we find that giving them this control means they are more willing to work through any challenges that follow.

\subsection{Group Projects $-$ Individual Grades}

Group work can often be unpopular, and some suggest they are particularly difficult to use in Computer Science unless great efforts are made to change the student culture \cite{WJDL04}.  By starting group projects on day one of university we hope to embed the view of teamwork as a normal part of CS education.  

One of the most common complaints about group work is that it can lead to an unfair distribution of grades with weaker team members being rewarded for the work of stronger members.  This is particularly acute at the start of a degree where students have very different levels of prior knowledge.  Although the projects are firmly team based the grades are individual, based on each student's contribution.  The grade is produced using a mixture of self and peer assessment moderated by a teacher, as described in the next section.  This follows a trend of similar approaches such as \cite{BM16}.

\section{Our Project Assessment Scheme}
\label{SEC:CA}

\subsection{Assessment Prior to 2018/19}

We previously graded projects based on a single final submission.  The bulk of the marks would be allocated individually according to a student's discussion of the product in a short ($\sim 10-15$ minute) viva.  The viva format works very well in ascertaining a student's understanding of the code and processes used, and we were confident in the integrity of the marks obtained. The final two weeks of project contact time were dedicated to the vivas, meaning students were not required outside their viva slot during these weeks.  Since their other classes would be gearing up for final exams and CW at this time it actually worked well to wind the projects down prior.

However, the end-point submission would often lead to little progress for much of the semester with a rush before the deadline.  This could lead to the strongest programmer doing the bulk of the work, and less focus paid to the team work and processes that the project was meant to emphasise.  We wanted to implement an assessment style that would promote steady work on the project throughout the semester: a continuous assessment approach. 

\subsection{Assessment from 2018/19 Onwards}

\begin{figure*}
\includegraphics[width=0.94\textwidth]{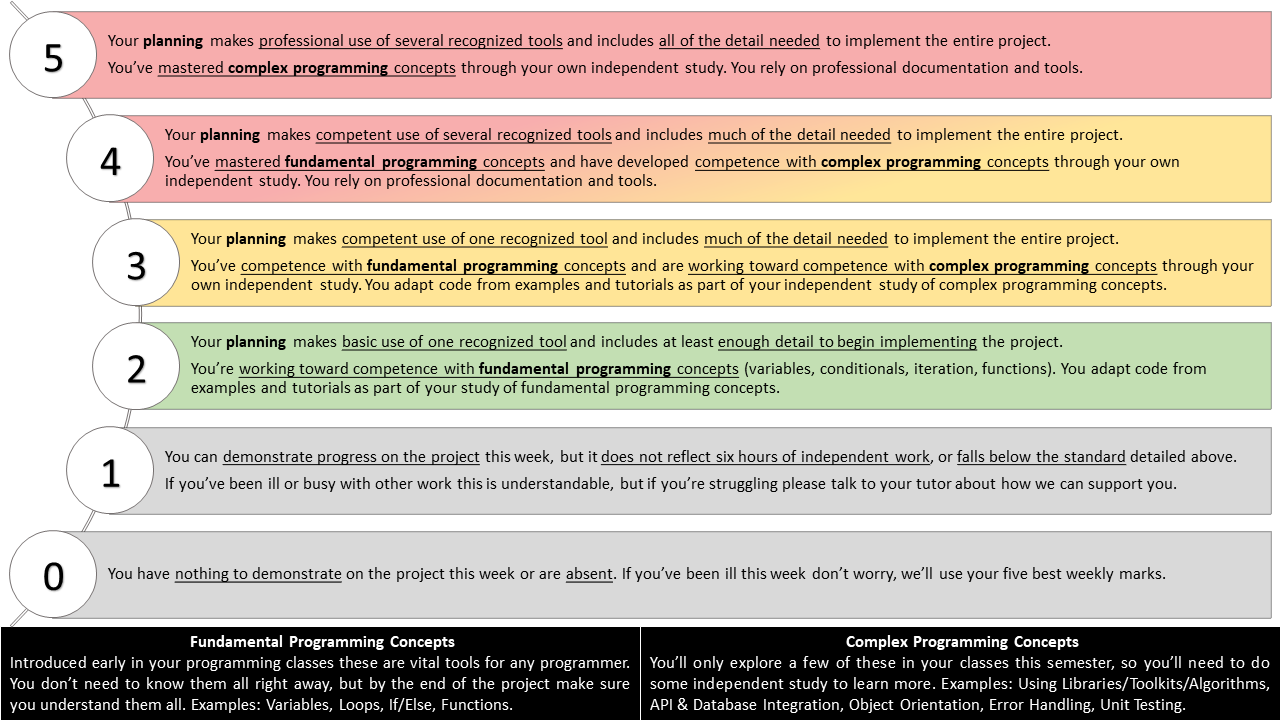}
\caption{Markscheme for weekly contribution in Semester 1 Project \label{fig:Grade}}
\Description[Markscheme for weekly contribution in Semester 1 Project]{Markscheme for weekly contribution in Semester 1 Project giving clear definitions on what we consider students need for the 6 different grades.}
\end{figure*}

Groups now give weekly presentations on their progress (members taking it in turn to present). This builds valuable public speaking / presentation skills.  Each student self-evaluates an individual mark for the week according to the rubric in Figure \ref{fig:Grade} which is included with a justification on the final slide.  There is no formal peer-grading but since it is a group presentation this prompts peer discussion and feedback $-$ the final choice of what number goes in the slide is made by the individual.  
The actual mark is decided by the teacher: if it differs from the student's suggestion then the rationale is explained openly.  There can be a debate but the final decision is the teacher's.  A student's module grade comes from:
\begin{itemize}
\item The average weekly mark, after discarding the lowest three (allows for late arrivals / illness without any make-up tasks).  
\item A viva at the end of the module, with questions about the final group code: to test individual student's understanding.
\item A piece of writing.  In Semester One this is an online professional portfolio\footnote{which they maintain throughout their degree to assist with applying for work.} and in Semester Two this is a piece of academic writing based on literature research.
\end{itemize}
The weekly grades are fairly consistent with the final viva grade but are on average higher since:  (1) weaker students can get credit for small contributions made to a greater whole which they couldn't explain alone; (2) exploration that does not work out and thus is not present in the final submission is rewarded; (3) students who spend time helping other team members or groups are rewarded for these efforts.  In each case we are happy to reward these contributions.   

\subsection{Small Changes from 2018/19 and 2019/20}
\label{SEC:ThisYear}

The assessment scheme above was introduced in academic year 2018/19 along with the new constitution of the projects as  modules.  In its first outing the group presentations took place individually, i.e. one group presented to the teacher and were graded at a time.  For 2019/20 we changed this so that other groups are in the audience for presentations.  Usually 7 groups will be in a room together.  

In the first outing there were student complaints about unfair grading. The module leader acted as a moderator, sitting in on each teacher's grading to ensure consistency.  However, the perception of unfairness proved hard to dispel. The decision was made for more open grading: by watching the presentations for other groups students gain a proper understanding of how others are working; while the open conversations around grading ensures everyone in the room understand the rationale behind a grade.  Although this makes no difference to actual marks, the trust it builds in the grading system is essential for student engagement, see also \cite{BM16}.


\section{Results}
\label{SEC:Results}

4006CEM saw a 34\% increase in attendance in 2018/19 compared to the Semester 1 CS ALL project the year before; and a 20\% increase in grades on the part of the assessment that was unchanged (the viva on the final code).  Similarly, in Semester 2 there was a 22\% increase in attendance and a 34\% increase in viva grades.  
Taking all modules together first year CS had 13\% higher attendance in 2018/19 than in 2017/18.  We cannot be sure whether this indicates good habits taking hold from the enforced attendance in ALL, or just a more studious cohort.  

Students are given evaluation questionnaires for each module,   but we cannot use these to directly measure the impact of continuous assessment:  since the old projects were not modules they did not receive a dedicated questionnaire.  
However, we note that in the 2018/19 questionnaires there was not a single negative comment relating to student groups.  This contrasts greatly with our prior experience where students would commonly complain that various team mates were not turning up or working hard enough.  

The negative questionnaire comments for 2018/19 were mostly on perceived inconsistent grading, which we addressed by the changes outlines in Section \ref{SEC:ThisYear}.  In the 2019/20 Semester 1 survey just taken there was not a single mention of inconsistency and satisfaction levels were considerably higher!

\section{Summary}

We have described our Year 1 activity-led CS projects:  
their course integrative nature which is highly valued by our university; 
continuous assessment which has improved attendance and reduced complaints about team members under-engaging; 
self-assessment \\ to give students skills in reflection and understanding mark schemes; 
and open grading to build trust and clarify our expectations.

\begin{acks}
The authors are grateful to all their colleagues who have taught and assessed for these project modules in the past and present. 
\end{acks}

\bibliographystyle{ACM-Reference-Format}
\bibliography{EdRes}


\begin{thebibliography}{9}


\ifx \showCODEN    \undefined \def \showCODEN     #1{\unskip}     \fi
\ifx \showDOI      \undefined \def \showDOI       #1{#1}\fi
\ifx \showISBNx    \undefined \def \showISBNx     #1{\unskip}     \fi
\ifx \showISBNxiii \undefined \def \showISBNxiii  #1{\unskip}     \fi
\ifx \showISSN     \undefined \def \showISSN      #1{\unskip}     \fi
\ifx \showLCCN     \undefined \def \showLCCN      #1{\unskip}     \fi
\ifx \shownote     \undefined \def \shownote      #1{#1}          \fi
\ifx \showarticletitle \undefined \def \showarticletitle #1{#1}   \fi
\ifx \showURL      \undefined \def \showURL       {\relax}        \fi
\providecommand\bibfield[2]{#2}
\providecommand\bibinfo[2]{#2}
\providecommand\natexlab[1]{#1}
\providecommand\showeprint[2][]{arXiv:#2}

\bibitem[\protect\citeauthoryear{Anderson, Peters, Halloran, Every,
  Shuttleworth, Liarokapis, Lane, and Richards}{Anderson et~al\mbox{.}}{2012}]%
        {APHESLLR12}
\bibfield{author}{\bibinfo{person}{E.F. Anderson}, \bibinfo{person}{C.E.
  Peters}, \bibinfo{person}{J. Halloran}, \bibinfo{person}{P. Every},
  \bibinfo{person}{J. Shuttleworth}, \bibinfo{person}{F. Liarokapis},
  \bibinfo{person}{R. Lane}, {and} \bibinfo{person}{M. Richards}.}
  \bibinfo{year}{2012}\natexlab{}.
\newblock \showarticletitle{In at the Deep End: {A}n Activity-Led Introduction
  to First Year Creative Computing}.
\newblock \bibinfo{journal}{\emph{Computer Graphics Forum}}
  \bibinfo{volume}{31}, \bibinfo{number}{6} (\bibinfo{year}{2012}),
  \bibinfo{pages}{1852--1866}.
\newblock
\urldef\tempurl%
\url{https://doi.org/10.1111/j.1467-8659.2012.03066.x}
\showURL{%
\tempurl}


\bibitem[\protect\citeauthoryear{Bergin and Mooney}{Bergin and Mooney}{2016}]%
        {BM16}
\bibfield{author}{\bibinfo{person}{Susan Bergin} {and} \bibinfo{person}{Aidan
  Mooney}.} \bibinfo{year}{2016}\natexlab{}.
\newblock \showarticletitle{An Innovative Approach to Improve Assessment of
  Group Based Projects}. In \bibinfo{booktitle}{\emph{Proc. Koli Calling
  2016}}. \bibinfo{publisher}{ACM}, \bibinfo{pages}{12--20}.
\newblock
\urldef\tempurl%
\url{https://doi.org/10.1145/2999541.2999543}
\showURL{%
\tempurl}


\bibitem[\protect\citeauthoryear{Blignaut and Venter}{Blignaut and
  Venter}{1998}]%
        {BV98}
\bibfield{author}{\bibinfo{person}{R.J. Blignaut} {and} \bibinfo{person}{I.M.
  Venter}.} \bibinfo{year}{1998}\natexlab{}.
\newblock \showarticletitle{Teamwork: can it equip university science students
  with more than rigid subject knowledge?}
\newblock \bibinfo{journal}{\emph{Computers \& Education}}
  \bibinfo{volume}{31}, \bibinfo{number}{3} (\bibinfo{year}{1998}),
  \bibinfo{pages}{265--279}.
\newblock
\urldef\tempurl%
\url{https://doi.org/10.1016/S0360-1315(98)00031-1}
\showURL{%
\tempurl}


\bibitem[\protect\citeauthoryear{Bowman, Jarratt, Culver, and Segre}{Bowman
  et~al\mbox{.}}{2019}]%
        {BJCS19}
\bibfield{author}{\bibinfo{person}{N.A. Bowman}, \bibinfo{person}{L. Jarratt},
  \bibinfo{person}{K.C. Culver}, {and} \bibinfo{person}{A.M. Segre}.}
  \bibinfo{year}{2019}\natexlab{}.
\newblock \showarticletitle{How Prior Programming Experience Affects Students'
  Pair Programming Experiences and Outcomes}. In
  \bibinfo{booktitle}{\emph{Proc. ITiCSE '19}}. \bibinfo{publisher}{ACM},
  \bibinfo{pages}{170--175}.
\newblock
\urldef\tempurl%
\url{https://doi.org/10.1145/3304221.3319781}
\showURL{%
\tempurl}


\bibitem[\protect\citeauthoryear{Croft and England}{Croft and England}{2019}]%
        {CE19}
\bibfield{author}{\bibinfo{person}{D. Croft} {and} \bibinfo{person}{M.
  England}.} \bibinfo{year}{2019}\natexlab{}.
\newblock \showarticletitle{Computing with {C}odio at {C}oventry {U}niversity:
  {O}nline Virtual {L}inux Boxes and Automated Formative Feedback}. In
  \bibinfo{booktitle}{\emph{Proc. CEP '19}}. \bibinfo{publisher}{ACM}, Article
  \bibinfo{articleno}{16}, \bibinfo{numpages}{4}~pages.
\newblock
\urldef\tempurl%
\url{https://doi.org/10.1145/3294016.3294018}
\showURL{%
\tempurl}


\bibitem[\protect\citeauthoryear{Croft and England}{Croft and England}{2020}]%
        {CE20}
\bibfield{author}{\bibinfo{person}{D. Croft} {and} \bibinfo{person}{M.
  England}.} \bibinfo{year}{2020}\natexlab{}.
\newblock \showarticletitle{Computing with {CodeRunner} at {C}oventry
  {U}niversity: {A}utomated summative assessment of {P}ython and {C}++ code}.
  In \bibinfo{booktitle}{\emph{Proc. CEP '20}}. \bibinfo{publisher}{ACM},
  \bibinfo{pages}{In Press}.
\newblock
\urldef\tempurl%
\url{https://doi.org/10.1145/3372356.3372357}
\showURL{%
\tempurl}


\bibitem[\protect\citeauthoryear{Hazzan, Lapidot, and Ragonis}{Hazzan
  et~al\mbox{.}}{2011}]%
        {HLR11}
\bibfield{author}{\bibinfo{person}{O. Hazzan}, \bibinfo{person}{T. Lapidot},
  {and} \bibinfo{person}{N. Ragonis}.} \bibinfo{year}{2011}\natexlab{}.
\newblock \bibinfo{booktitle}{\emph{Guide to Teaching Computer Science: An
  Activity-Based Approach}}.
\newblock \bibinfo{publisher}{Springer}.
\newblock


\bibitem[\protect\citeauthoryear{Vihavainen, Paksula, and
  Luukkainen}{Vihavainen et~al\mbox{.}}{2011}]%
        {VPL11}
\bibfield{author}{\bibinfo{person}{A. Vihavainen}, \bibinfo{person}{M.
  Paksula}, {and} \bibinfo{person}{M. Luukkainen}.}
  \bibinfo{year}{2011}\natexlab{}.
\newblock \showarticletitle{Extreme Apprenticeship Method in Teaching
  Programming for Beginners}. In \bibinfo{booktitle}{\emph{Proc. SIGCSE '11}}.
  \bibinfo{publisher}{ACM}, \bibinfo{pages}{93--98}.
\newblock
\urldef\tempurl%
\url{http://doi.org/10.1145/1953163.1953196}
\showURL{%
\tempurl}


\bibitem[\protect\citeauthoryear{Waite, Jackson, Diwan, and Leonardi}{Waite
  et~al\mbox{.}}{2004}]%
        {WJDL04}
\bibfield{author}{\bibinfo{person}{W.M. Waite}, \bibinfo{person}{M.H. Jackson},
  \bibinfo{person}{A. Diwan}, {and} \bibinfo{person}{P.M. Leonardi}.}
  \bibinfo{year}{2004}\natexlab{}.
\newblock \showarticletitle{Student Culture vs Group Work in Computer Science}.
\newblock \bibinfo{journal}{\emph{SIGCSE Bull.}} \bibinfo{volume}{36},
  \bibinfo{number}{1} (\bibinfo{year}{2004}), \bibinfo{pages}{12--16}.
\newblock
\urldef\tempurl%
\url{http://doi.org/10.1145/1028174.971308}
\showURL{%
\tempurl}


\end{thebibliography}

\end{document}